\documentclass[aps,prc,twocolumn,amsmath,superscriptaddress,floatfix,nofootinbib]{revtex4-1}

\usepackage{setspace,ulem, epsfig,amssymb,amsfonts,amsmath,mathtools,bm,color,xcolor,graphicx,braket,adjustbox,esint,upgreek, verbatim,ctable}

\usepackage[colorlinks, linkcolor=red,anchorcolor=green,citecolor=blue]{hyperref}

\begin{document}

\title{The Gilbert Damping Factor of Heavy Quark Spin Polarization in the Magnetic Field}

\author{Tianyang Li}
\affiliation{Department of Physics, Tianjin University, Tianjin 300350, China}

\author{Anping Huang}\email{huanganping425@cumt.edu.cn}
\affiliation{School of Material Science and Physics, China University of Mining and Technology, Xuzhou 221116, China}

\author{Baoyi Chen}
\email{baoyi.chen@tju.edu.cn}
\affiliation{Department of Physics, Tianjin University, Tianjin 300350, China}

\date{\today}
\begin{abstract}

We employ the linear response theory to calculate the polarization rate of heavy quark spin in the presence of a strong magnetic field and the hot QCD matter, both of which are simultaneously generated in relativistic heavy-ion collisions. The hot QCD medium is simplified as a fermionic system consisting of only quarks. The spin of heavy quarks can be polarized as a result of combined contributions from spin-spin interactions between quarks and spin-magnetic field interactions. This spin dynamics is modeled as consisting of a polarization term and a dissipation term, which is described by the Landau-Lifshitz-Gilbert (LLG) equation and widely studied in condensed matter physics, analogous to the momentum evolution in the Langevin equation. In this study, we calculate the Gilbert damping factor that characterizes the spin polarization rate of heavy quarks, considering a Coulomb potential between two fermions in the medium. The dependence of the heavy quark spin polarization rate on the strength of the magnetic field, the heavy quark mass, temperature, and baryon chemical potential is studied in detail. This analysis contributes to a better understanding of quark spin dynamics in the hot QCD medium and the magnetic field.

\end{abstract}

\maketitle
\noindent

\section{Introduction} 
Over the past few decades, extensive studies have investigated the strong interactions under extreme conditions in Relativistic Heavy-Ion Collisions (RHIC) and at the Large Hadron Collider (LHC). These heavy-ion collisions produce a new form of deconfined matter, where the degrees of freedom are partons rather than hadrons, through a deconfinement phase transition~\cite{STAR:2005gfr,Bazavov:2011nk}. This phase transition becomes a crossover at low baryon chemical potentials and a first-order transition at high baryon chemical potentials. The abnormal yield suppression of heavy quarkonium was proposed over forty years ago to probe the signals of this new deconfined matter~\cite{Matsui:1986dk}. Since then, various observables have been introduced to focus on different properties of the hot QCD matter, including thermal photon spectrum~\cite{Paquet:2015lta}, collective flows of light hadrons~\cite{Shen:2014vra}, heavy-quark energy loss~\cite{He:2012df}, quarkonium dissociation and regeneration~\cite{Yan:2006ve}, jet quenching~\cite{Cao:2020wlm}, and more.

In addition to the hot deconfined medium, strong electromagnetic fields are generated by the fast moving spectator protons in semi-central nuclear collisions~\cite{Deng:2012pc}. These electromagnetic fields produce quark-antiquark pairs and dileptons~\cite{Tuchin:2013ie}, particularly in the peripheral collisions. This phenomenon has been observed in the enhancement of the nuclear modification factor $R_{AA}$ of heavy quarkonium in the extremely low transverse momentum region in peripheral Pb-Pb collisions at $5.02$ TeV~\cite{Shi:2017qep}. Furthermore, the electromagnetic fields also affect the momentum and spin orientation of quarks in the hot QCD medium, leading to the generation of direct flows~\cite{STAR:2014clz,Zhang:2022lje} as well as spin polarization of light quarks and the corresponding light hadrons~\cite{Guo:2019joy,Liu:2024hii}.

Even though the rotation of the medium leads to collective flows and spin polarization of quarks~\cite{Liang:2004ph,STAR:2017ckg}, the experimental data about $D$ and $\bar{D}$ (and $\Lambda$ and $\bar{\Lambda}$) suggest that the contribution of the magnetic field cannot be ignored~\cite{Liu:2024hii,Becattini:2016gvu}, which gives opposite modifications on charm and anti-charm quarks. The lifetime of the magnetic field is roughly estimated as $2R_A/(\gamma_L)$, where $R_A$ is the nuclear radius, and $\gamma_L$ is the Lorentz factor. Heavy quarks are produced in the initial parton hard scatterings within a very short time scale, and subsequent thermal production is significantly suppressed because of their large mass. Consequently, magnetic fields may modify both the momentum and spin orientation of heavy quarks. Furthermore, the hot QCD medium, consisting of charged fermions, helps to enhance the lifetime and strength of in-medium magnetic fields~\cite{Huang:2017tsq,Jiang:2022uoe}. Therefore, it is essential to study the effects of magnetic fields on the dynamical evolution of heavy quarks.

On the other hand, experiments have measured the spin alignment of $J/\psi$ in relativistic heavy-ion collisions~\cite{ALICE:2022dyy}. $J/\psi$ originates from both primordial production and regeneration via the coalescence of charm and anti-charm quarks in the hot QCD matter. The collective behaviors and spin polarization of heavy quarks are inherited by the regenerated $J/\psi$. Previous theoretical calculations based on the Boltzmann transport model have computed the $J/\psi$ spin alignment~\cite{Zhao:2023plc}, considering two extreme cases: no polarization and complete polarization of the charm quark spins. To incorporate realistic spin polarization of charm quarks, both spin-magnetic field interactions and spin-angular momentum coupling must be considered simultaneously. The Landau–Lifshitz–Gilbert (LLG) equation has been employed to study the spin polarization of fermions in the magnetic field and the condensed matter physics~\cite{PhysRevLett.102.086601}. It is also extended to investigate the spin dynamics of heavy quarks in heavy-ion collisions~\cite{Liu:2024hii}. A critical parameter in this context is the Gilbert damping factor, which characterizes the rate of heavy quark spin polarization. In this work, we will calculate its value based on linear response theory.

This work is organized as follows. Section II introduces the LLG equation and the parameters inside. Section III derives the formula for the Gilbert damping factor, which represents the polarization rate of the heavy quark. In Section IV, numerical results for the damping factor are provided under different quark masses, temperatures, and baryon chemical potentials. Finally, Section V gives a summary.

\section{theoretical model}
\subsection{LANDAU–LIFSHITZ–GILBERT EQUATION}
\label{sec:llg}
In the presence of magnetic fields and a hot medium composed of partons, the spin of heavy quarks undergoes precession around the direction of the magnetic field, while simultaneously generating polarization. This polarization results from the combined effects of spin-spin interactions between the heavy quark and light quarks, as well as spin-magnetic field interactions. Consequently, the spin of the heavy quark evolves toward the direction of the magnetic field, \( \mathbf{B} \). At the same time, random spin-spin interactions cause the spin to align in various directions. This effect can be parameterized by a randomly fluctuating term, \( \mathbf{B}_{\rm th} \). A detailed balance exists between these two processes, which governs the degree of heavy quark spin polarization once equilibrium is reached. Therefore, the dynamical equation governing the evolution of the heavy quark spin can be expressed as follows:
~\cite{PhysRevB.71.174430}, 
\begin{align}
    \frac{d{\bf s}}{dt}=-{|\gamma|} {\bf s}\times ({\bf B}+{\bf B}_{\rm th}) -{\alpha }{\bf s}\times \frac{d{\bf s}}{dt}.
    \label{eq-llg}
\end{align}
Here, \( \mathbf{s} = \mathbf{S}/|\mathbf{S}| \) is a unit vector, where \( \mathbf{\mu} = g \mathbf{J}_s q_e / (2m_q) \) represents the magnetic moment. \( |\mathbf{J}_s| = 1/2 \) is the spin of the heavy quark, and \( q_e \) and \( m_q \) are the electric charge and mass of the quark, respectively. The parameter \( \gamma \) is defined as \( \gamma = \mu / J_s = g q_e / (2m_q) \), with the constant factor \( g = 2 \)~\cite{PhysRevD.46.3529}. Spin polarization is primarily induced by the second term, where \( \alpha \) is understood to be the Gilbert damping factor. By making a first-order approximation of \( \frac{d\mathbf{s}}{dt} \) on the right-hand side of Eq.(\ref{eq-llg}), we can easily obtain the Landau–Lifshitz–Gilbert (LLG) equation, which can be solved numerically. In the following sections, we will focus exclusively on the Gilbert damping factor, which characterizes the polarization rate of the fermions. The dissipation term represented by the noise term \( \mathbf{B}_{\rm th} \)~\cite{PhysRevB.83.054432} is neglected here. The Gilbert damping factor, \( \alpha \), has been calculated using linear response theory~\cite{kambersky1976ferromagnetic, PhysRevB.76.134416,PhysRevB.90.014420}, which is also related to the Hamiltonian of the medium. Both approaches will be discussed in detail to derive the formula for \( \alpha \).

\subsection{Hamiltonian of the fermionic system\label{subsec:The-method-to}}
The evolution of fermion spin in magnetic fields and within a medium has been studied using the theory of spin waves in ferromagnets~\cite{PhysRev.174.227}. Assuming that the magnetic field is directed along the negative \( z \)-axis, the normalized vector of the heavy quark spin can be expressed as follows, taking into account the precession process,
\begin{equation}
{\bf s}=(ce^{i(qx-\omega t)},de^{i(qx-\omega t)},s_z)+c.c.,
\label{eq-s-wave}
\end{equation}
Here, c.c. denotes the complex conjugate, and \( q \) represents the spin wave vector. In the limit of complete polarization, \( s_z \) simplifies to \( s_z \approx 1 \), independent of both \( \omega \) and \( t \), with \( c \) and \( d \) becoming small values, well below unity. In the case of pure precession, the frequency \( \omega \) is real. However, when considering spin polarization, \( \omega(q) = \omega_1 - i\omega_2 \) becomes complex. Here, \( \omega_1 = \gamma B_{ext} \) characterizes the precession frequency~\cite{edwards2009quantum}, where \( B_{ext} \) is the strength of the external magnetic field. The term \( \omega_2 \) represents the decay of the transverse component of the spin vector.

By substituting the spin wave equation (Eq.~\ref{eq-s-wave}) into the LLG equation (Eq.~\ref{eq-llg}), we obtain the relationship between the damping factor \( \alpha \) and the complex frequency \( \omega \). Assuming that \( \alpha \) is small, the term \( 1 + \alpha^2 \) in the denominator of the LLG equation can be approximated as unity. Consequently, the damping factor in Eq.~\ref{eq-llg} is given by~\cite{PhysRevB.90.014420} (for details, see Appendix~\ref{sec:appendix-w2}).
\begin{align}
    \alpha = \frac{\omega_2}{\omega_1}.
    \label{eq-alpha-w2}
\end{align}
We will derive the expression for \( \omega_2 \) and the corresponding \( \alpha \) in the following sections. To determine the imaginary part of the frequency in the spin vector, the Hamiltonian of the fermionic system, considering the presence of the external magnetic field, can be constructed as~\cite{kambersky1976ferromagnetic}.
\begin{align}
    H = H_{ k} + H_{ext}+ H_{ee} + H_Z + H_{SO}
    \label{eq-hamil}
\end{align}
where \( H_{ k} \) and \( H_{ ext} \) represent the kinetic energy and the potential energy of the fermions in the external field, respectively. \( H_{ee} \) describes the fermion-fermion interactions within the system. \( H_Z \) and \( H_{ SO} \) account for the energy shifts due to the Zeeman effect and spin-orbit coupling, respectively. In a system consisting of \( n \) fermions, the energy shifts can be expressed as follows,
\begin{align}
H_{\mathrm{Z}}&=-2\frac{\mu_{Bo}}{\hbar }B_{{ext}}S^z_{tot}-\frac{\mu_{Bo}}{\hbar }B_{{ext}}L_{{tot}}^{z},\\
H_{SO}&=\sum_{i=1}^{n}\xi\mathbf{L}_{i}\cdot\mathbf{S}_{i}.
\end{align}
\( S^z_{ tot} \) denotes the \( z \)-component of the spin for all fermions. $\mu_{Bo}$ is the Bohr magneton. The \( z \)-direction is chosen to align with the direction of the external magnetic field, \( {\bf B} = (0, 0, -B_{ ext}) \). \( L_z^{tot} \) represents the \( z \)-component of the total angular momentum of the fermionic system. \( {\bf L}_i \) and \( {\bf S}_i \) denote the angular momentum and spin of the \( i \)-th fermion in the system, respectively. Here, \( \xi \) represents the strength of the spin-orbit coupling, which can be calculated by $\xi=\frac{1}{2 m^{2} c^{2} r} \frac{\mathrm{~d} U}{\mathrm{~d} r}$\cite{PhysRevB.79.094422} . We provide further discussion in Appendix \ref{spin-orbit}. 
Given the Hamiltonian of the fermionic system, the time evolution of the spin operator \( {\bf S} \) in the Heisenberg picture is governed by the following equation,
\begin{align}
    \frac{d{\bf S}(t)}{dt} = \frac{1}{i\hbar}[{\bf S}(t), H]
    \label{eq-h-system}
\end{align}
Using the dynamical equation in Eq.(\ref{eq-h-system}) and the spin wave formula in Eq.(\ref{eq-s-wave}), one can calculate the imaginary part of the frequency \( \omega_2 \), which is related to the Gilbert damping factor via Eq.(\ref{eq-alpha-w2}).

\subsection{Gilbert damping factor}
\label{subsec:Calculate-the-Gilbert-damping}
When analyzed in the frequency domain , the transverse magnetization ${\bf M}_\perp = (M_x, M_y, 0)$ of the fermionic system can be calculated by introducing an effective perturbative transverse magnetic field ${\bf B}_\perp({\bf r}, t)$ in momentum space:
\begin{align}
    \mathbf{M_{\bot}}(\mathbf{q}, \omega') = \chi_{\bot}(\mathbf{q}, \omega') \mathbf{B_{\bot}}(\mathbf{q}, \omega').
\end{align}
Here, $\chi_\perp = \chi_{-+}$ denotes the transverse magnetic susceptibility of the fermionic system. Within the framework of linear response theory, $\chi_{-+}$ is related to the spin via,
\begin{align}
    \chi_{-+}(\boldsymbol{q},\omega')=\int dt\left\langle \left\langle S_{\boldsymbol{q}}^{-}(t),S_{-\boldsymbol{q}}^{+}(0)\right\rangle \right\rangle e^{-i\omega'_{-}t},
\end{align}
where $\omega'_-=\omega'-i\eta$, with $\eta$ to be an infinitesimal number. The spin term of the entire system is $S^{\pm}=\sum_{i}e^{i\mathbf{q}\cdot\mathbf{r_{i}}}S_{i}^{\pm}$, where ${\bf r}_i$ is the position of the $i$-th fermion, $S_i^{\pm}$ is the raising and lowering operator of the spin for the $i$-th particle. The sum covers all the fermions in the system.   The term $\left\langle \left\langle S_{\boldsymbol{q}}^{-}(t),S_{-\boldsymbol{q}}^{+}(0)\right\rangle \right\rangle $ is defined as
$\frac{i}{\hbar}\theta\left(t-t^{\prime}\right)\left\langle \left[S^{+}(t),S^{-}\left(t^{\prime}\right)\right]\right\rangle $, and $t$ and $t'$ are the time points. 
The transverse magnetic susceptibility $\chi_{-+}$ can be further calculated as,
 \begin{align}
   \label{eq-chi-simp2}
     \chi_{-+}(\boldsymbol{q},\omega') & =
\frac{-1}{\hbar\omega'_{-}}(-\left\langle \left[S_{\boldsymbol{q}}^{-},S_{-\boldsymbol{q}}^{+}\right]\right\rangle\nonumber \\
&  -\int_{0}^{\infty}dt\left\langle \left[\frac{dS_{\boldsymbol{q}}^{-}(t)}{dt},S_{-\boldsymbol{q}}^{+}\right]\right\rangle e^{-i\omega'_{-}t}).
\end{align}

In the Heisenberg picture, take the formula of the spin wave into the Heisenberg equation, one can obtain 
\begin{align}
    i\hbar \left\langle  \frac{\partial S_{\boldsymbol{q}}^{-}(t)}{\partial t}\right\rangle =-\hbar(\omega_1+i\omega_2)\left\langle S_{\mathbf{\boldsymbol{q}}}^{-}(t)\right\rangle.
    \label{eq-s-evolve}
\end{align}
Perform the average of the operator which will be used in the $\chi_{-+}({\bf q},\omega')$, we have
\begin{align}
\label{eq-trace-S}
    \frac{\partial\left\langle S_{\boldsymbol{q}}^{-}(t)\right\rangle }{\partial t}&=\frac{\partial\mathrm{tr(\rho}S_{\boldsymbol{q}}^{-}(t))}{\partial t}
=\left\langle \frac{\text{\ensuremath{\partial}}S_{\boldsymbol{q}}^{-}(t)}{\partial t}\right\rangle 
\end{align}
and the corresponding transverse magnetic susceptibility becomes,
\begin{align}
\chi_{-+}(\boldsymbol{q},\omega')=\frac{1}{\hbar\omega'_{-}}\text{\ensuremath{\left(\left\langle \left[S_{\boldsymbol{q}}^{-},S_{-\boldsymbol{q}}^{+}\right]\right\rangle +\text{\ensuremath{\hbar(\omega_1+i\omega_2)}\ensuremath{\chi_{-+}}(\ensuremath{\boldsymbol{q}},\ensuremath{\omega'})}\right)}}.
\end{align}
Considering the commutation relation of the total spin in the fermionic system consisting of $n$ fermions, 
\begin{align}\left[S_{\boldsymbol{q}}^{-},S_{-\boldsymbol{q}}^{+}\right] & =-2\hbar\sum_{n}S_{n}^{z}=-2\hbar S_{z},
\end{align}
In the case close to complete spin polarization, the $S_{n}^{z}$ is taken as $S_{n}^{z}=1/2$. A simple relation between the transverse magnetic susceptibility and the complex frequency of the spin wave is obtained, 
\begin{equation}
\chi_{-+}(\omega')=-\frac{2\left\langle S_{z}\right\rangle }{(\omega'-\omega_1-i\omega_2)}.
\label{eq-chi-final-1}
\end{equation}

In the other approach to calculate $\chi_{-+}(\omega')$, the term 
$\frac{dS_q^-(t)}{dt}$ in the Eq.(\ref{eq-chi-simp2}) of the transverse magnetic susceptibility can be directly calculated with the Heisenberg equation Eq.(\ref{eq-h-system}). The corresponding $\chi_{-+}$ therefore becomes~\cite{PhysRevB.90.014420}, 
\begin{align}
\label{eq-chi-final-2-simplified}
\chi_{-+}(\omega')&=\frac{-2\left\langle S_{z}\right\rangle }{\left(\omega'-\omega_{1}\right)} +\frac{ \chi_{A}(\omega')-\left\langle \left[A^{-},S^{+}\right]\right\rangle }{\hbar^{2}\left(\omega'-\omega_{1}\right)^{2}},
\end{align}
with the definition of $\chi_A(\omega')$,
\begin{align}
\label{eq-chi-A}
\chi_A(\omega') &= \frac{i}{\hbar}\int_{0}^{\infty}dt\left\langle \left[A^{-}(t),A^{+}\right]\right\rangle e^{-i\omega'_{-}t}.
\end{align}
In the above equation, $A^{-}\equiv \left[S^{-},H_{\mathrm{SO}}\right]$. For the details of derivation, please see the Appendix \ref{sec:appendix-b.-Calculate}.

Both Eq.(\ref{eq-chi-final-1}) and Eq.(\ref{eq-chi-final-2-simplified}) are the formula of $\chi_{-+}(\omega')$ calculated with two different methods. To connect the imaginary part of the spin frequency $\omega_2$ with the Hamiltonian of the system and $\chi_A(\omega')$, we perform a expansion in Eq.(\ref{eq-chi-final-1}), 
\begin{align}\chi_{-+}(\omega') & =\frac{-2\left\langle S_{z}\right\rangle }{\left(\omega'-\omega_{1}-\Delta\omega-i\omega_{2}\right)}
\nonumber \\
\label{eq-chi-final-1-simplified}
 & \thickapprox\frac{-2\left\langle S_{z}\right\rangle }{\left(\omega'-\omega_{1}\right)}-\frac{2\left\langle S_{z}\right\rangle }{\left(\omega'-\omega_{1}\right)^{2}}\left(\Delta\omega+i\omega_{2}\right).
\end{align}
where $\Delta\omega$ represents a small precession frequency shift caused
by the spin-orbit coupling term. Comparing two formula about $\chi_{-+}(\omega')$ in Eq.(\ref{eq-chi-final-1-simplified}) and Eq.(\ref{eq-chi-final-2-simplified}), the expression about $\omega_2$ is obtained, 
\begin{equation}
\omega_{2}=-\frac{1}{2\hbar^{2}\left\langle S_{z}\right\rangle }\mathrm{Im}\chi_{A}\left(\omega'\right)
\label{eq-w2-Im}
\end{equation}
during the derivation the relation $\mathrm{Im}\left\langle \left[A^{-},S^{+}\right]\right\rangle=0 $ is employed~\cite{PhysRevB.90.014420}. 
To connect with the Hamiltonian of the fermionic system, we rewrite the $\chi_{A}\left(\omega'\right)$ as a function of the density matrix,
\begin{align}
\chi_{A}\left(\omega'\right)&=\frac{i}{\hbar}\int_{0}^{\infty}dt\mathrm{\mathit{Tr}\left(\mathit{\hat{\rho}\left[A_{\boldsymbol{q}}^{-}(t),A_{\boldsymbol{q}}^{+}\right]}\right)}e^{-i\omega'_{-}t}
\end{align}
In the grand canonical ensemble, the density matrix is $\hat{\rho}=\frac{1}{\Xi}\exp(-\alpha'\hat{N}-\beta\hat{H})$ with $\alpha'=-\mu_B/k_bT$, $\beta=1/k_bT$. $\mu_B$ is chemical potential and $T$ is temperature. $k_b$ is Boltzmann constant.  
With the second quantization, the above expression of $\chi_A(\omega)$ is calculated to be, 
\begin{align}
\chi_{A}(\omega') &= \frac{i}{\hbar} \int_{0}^{\infty} dt \, \sum_{f,i} e^{i (\epsilon_{i} - \epsilon_{f}) t / \hbar} \left| A_{fi}^{-} \right|^{2} \nonumber \\
&\qquad\qquad \times \mathrm{Tr} \left( \hat{\rho} \left( \hat{n}_{f} - \hat{n}_{i} \right) \right) e^{-i \omega'_{-} t},
\label{eq:22}
\end{align}
where $\hat{n}_{i,f}$ are the operator of the particle number in the initial state and the final state, respectively. $\epsilon_{i,f}$ are the eigenvalues of the initial and final states of the Hamiltonian $H_{single}$, respectively. $H_{single}$ represents the Hamiltonian
of a single particle in the system.
In the grand canonical ensemble, the trace term in above equation satisfies
\begin{equation}
Tr\left(\hat{\rho}\left(\hat{n}_{f}-\hat{n}_{i}\right)\right)=f(\epsilon_{f})-f(\epsilon_{i}),
\label{eq-23-trace}
\end{equation}
where $f(\epsilon)$ represents the Fermi-Dirac distribution, and $\epsilon$ is the energy. For the detailed derivation, please see the Appendix~\ref{lab-append-c}

The imaginary part $\omega_2$ of the complex frequency then becomes,
\begin{align}
\omega_{2} &=-\frac{1}{2\hbar^{2}\left\langle S_{z}\right\rangle }\mathrm{Im}\chi_{A}\left(\omega'\right)
\nonumber\\ &=\frac{\pi}{2\hbar^{2}\left\langle S_{z}\right\rangle }\mathrm{\mathit{\sum_{f,i}\left|A_{fi}^{-}\right|^{2}\left(f\mathrm{(}\epsilon_{i}\mathrm{)}-f\mathrm{(}\epsilon_{f}\mathrm{)}\right)\delta\left(\hbar\omega'-\epsilon_{f}+\epsilon_{i}\right)},}
\end{align}
and the corresponding Gilbert damping factor is 
\begin{align}
\alpha&=\frac{\pi}{2\hbar^{2}\left\langle S_{z}\right\rangle \omega_{1}}\mathrm{\mathit{\sum_{f,i}\left|A_{fi}^{-}\right|^{2}\left(f\mathrm{(}\epsilon_{i}\mathrm{)}-f\mathrm{(}\epsilon_{f}\mathrm{)}\right)\delta\left(\hbar\omega'-\epsilon_{f}+\epsilon_{i}\right)}.}
\end{align}
The damping factor $\alpha$ describes the polarization process of spin in an external field, which is closely related to the total Hamiltonian $H=H_k + H_{ext}+H_{ ee}+H_Z+H_{SO}$. Due to the inclusion of the operator $A^-$ , $\alpha$ is primarily dominated by spin-orbit coupling, but the interactions between particles and the effects of the external field are also crucial for calculating $\alpha $. In the next section, we will perform calculations in this regard.

\section{Gilbert damping factor with scatterings}
\subsection{Gilbert damping factor with an external field}
 This section will introduce an external field $V(r)$ in the Hamiltonian of the fermionic system to derive the Gilbert damping factor for the spin evolution of the fermions. 
The expression of $\chi_A(\omega')$ is rewritten with the time evolution operator $U(t,t_0)$ where the terms of particle scatterings can be encoded,
\begin{align}
\label{eq-chi-A-scatter}
\chi_{A}(\omega') 
&= \frac{i}{\hbar} \int_{0}^{\infty} dt\,
\sum_{f,i} 
\left\langle f \left| U^{\dagger}(t,t_{0}) A^{-} U(t,t_{0}) \right| i \right\rangle \nonumber \\
& \qquad \times
\left\langle i \left| A^{+} \right| f \right\rangle 
\left( f(\epsilon_{f}) - f(\epsilon_{i}) \right) 
e^{-i\omega'_{-}t}.
\end{align}
To account for particle scatterings in the time evolution operator, a spherically symmetric potential $V(r)$ is introduced, and the Hamiltonian of a single particle is $H=H_k +V(r)$, where $H_k$ is the kinetic term. Solve the differential equation for the time evolution operator 
$i\hbar\frac{d}{dt}U_{I}\left(t,t_{0}\right)=V_{I}(t)U_{I}\left(t,t_{0}\right)$, with the solution, one obtains the matrix elements used in the $\chi_A(\omega')$, 
\begin{equation}
\left\langle f\left|U_{I}\left(t,t_{0}\right)\right|i\right\rangle =\delta_{fi}-\frac{i}{\hbar}\langle f|T|i\rangle\int_{t_{0}}^{t}e^{i\omega_{fi}t-\epsilon t}dt^{\prime}\label{eq-29}
\end{equation}
where $\epsilon$ is a small positive number. According to the Born approximation,
the $T$ operator in above equation satisfies $T=V+V\frac{1}{E-H_{0}+i\epsilon}V+V\frac{1}{E-H_{0}+i\epsilon}V\frac{1}{E-H_{0}+i\epsilon}V+\cdots$
and is $T\approx V$ in the first-order approximation.
The matrix of the $A^-$ in Eq.(\ref{eq-chi-A-scatter}) is written as 
\begin{equation}
A^{-}=\frac{\xi}{2}\begin{bmatrix}L^{-} & 0\\
2L^{z} & -L^{-}
\end{bmatrix}.
\end{equation}
with $L^{-}=L^{x}-iL^{y}$ to be the lowering operator of the angular momentum. $\xi$ is the spin-orbital coupling strength in $H_{SO}$ takes the form of  $\xi=\frac{1}{2 m^{2} c^{2} r} \frac{\mathrm{~d} U}{\mathrm{~d} r}$ and $U$ is external potential. In the scatterings with a central potential $V(r)$, we choose a spherical basis for the calculation of the elements of the T-matrix, the eigenstates of the spherical wave is labeled with the energy $E$, angular momentum $l$ and the magnetic quantum number $m$,
\begin{equation}
\langle E,l,m|T|E',l',m'\rangle=T_{l}(E)\delta_{ll'}\delta_{mm'}.
\end{equation}
The transition element $T_l(E)$ is related to the phase shift $\delta_l(E)$ in the scattering process. The corresponding total cross section $\sigma^{tot}$ will be calculated via the partial wave analysis, 
\begin{align}
T_{l}(E)&=-\frac{e^{i\delta_{l}(E)}\sin\delta_{l}(E)}{\pi},\label{eq:32} 
\\
\sigma^{tot}&=\frac{4\pi}{k^{2}}\sum_{l=0}^{\infty}(2l+1)\sin^{2}\delta_{l},
\end{align} 
where $p=\hbar k$ represents the magnitude of the particle's momentum. 
Therefore, the scattering cross section of the partial wave with a angular momentum $l$ is related to the phase shift as,
\begin{equation}
\label{eq-part-wa}
\sigma_{l}=\frac{4\pi}{k^{2}}(2l+1)\sin^{2}\delta_{l}\approx\frac{4\pi}{k^{2}}(2l+1)\delta^2_{l}.
\end{equation}
The approximation in above equation is employed when the phase shift $\delta_l$ is small. 

In a system with only an external central potential $V(r)$ and without the magnetic field, the Gilbert damping factor is related to the scattering cross-section of the partial wave via Eq.(\ref{eq-chi-A-scatter}-\ref{eq-part-wa}), and the imaginary part of $\chi_A(\omega')$ is,
\begin{align}
\mathrm{Im}\chi_{A}(\omega') 
=& \sum_{f,i,a,b} (f(\epsilon_{f}) -f(\epsilon_{i}))\bigg(
 A_{fi}^{-} \pi \delta(\omega' - \omega_{f}+\omega_i) \nonumber\\
& + \frac{1}{\hbar} \frac{\delta_{l}^{2}(\epsilon_{f})}{\pi} 
  A_{ab}^{-} \delta_{bi} \frac{1}{(\omega' - \omega_{f}+\omega_i)^{2}} \nonumber \\
& -\frac{1}{\hbar} \frac{\delta_{l}(\epsilon_{f})}{\pi} 
  A_{ab}^{-} \delta_{bi} 
  \frac{ \pi \delta\left( \frac{(\omega' - \omega_{f}+\omega_i)^{2}}{2\omega'\omega_{1}} \right) }{2\omega'(\omega_f-\omega_i)} \nonumber \\
& + \frac{1}{\hbar} \delta_{fa} A_{ab}^{-} 
  \frac{\delta_{l}^{2}(\epsilon_{i})}{\pi} 
  \frac{1}{(\omega' - \omega_{f}+\omega_i)^{2}} \nonumber\\
& + \frac{1}{\hbar} \delta_{fa} A_{ab}^{-} 
  \frac{\delta_{l}(\epsilon_{i})}{\pi} 
  \frac{ \pi \delta\left( \frac{(\omega' - \omega_{f}+\omega_i)^{2}}{2\omega'\omega_{1}} \right) }{2\omega'(\omega_f-\omega_i)} 
\bigg) A_{if}^{+}
\end{align}
where $\eta\mathrm{(}\epsilon\mathrm{)}$ is the derivative of
$f\mathrm{(}\epsilon\mathrm{)}$ with respect to energy $\epsilon$ and $\epsilon_{i,f}=\hbar \omega_{i,f}$. $\omega'$ is understood as the frequency of the transverse magnetic field. As without the contribution of the magnetic field in $\chi_A(\omega')$, $\mathrm{Im}\chi_A(\omega')$ becomes zero. 
Then we consider the effect of magnetic field in Gilbert damping factor and the Hamiltonian of the fermionic system becomes, 
\begin{equation}
H'=H_{k}+V(r)-\gamma S_{z}B_{ext}.
\end{equation}
{$H_{k}$ is the kinetic energy, and $\gamma$ is the gyromagnetic ratio.} Taking this into $A^-$, the transition rate from the state $i$ with spin $-\frac{1}{2}$ to the state $f$ with spin $+\frac{1}{2}$ becomes non-zero anymore. 
The formula of $\mathrm{Im}\chi_{A}\left(\omega'\right)$ after considering the magnetic field is
\begin{align}
\label{eq-chiA-final-sigma}
\mathrm{Im}\chi_{A}(\omega') 
=& \hbar \omega_{1}\sum_{f,i,a,b} \eta(\epsilon_{f}) \bigg(
 A_{fi}^{-} \pi \delta(\omega' - \omega_{1}) \nonumber\\
& + \frac{1}{\hbar} \frac{\delta_{l}^{2}(\epsilon_{f})}{\pi} 
  A_{ab}^{-} \delta_{bi} \frac{1}{(\omega' - \omega_{1})^{2}} \nonumber \\
& -\frac{1}{\hbar} \frac{\delta_{l}(\epsilon_{f})}{\pi} 
  A_{ab}^{-} \delta_{bi} 
  \frac{ \pi \delta\left( \frac{(\omega' - \omega_{1})^{2}}{2\omega'\omega_{1}} \right) }{2\omega'\omega_{1}} \nonumber \\
& + \frac{1}{\hbar} \delta_{fa} A_{ab}^{-} 
  \frac{\delta_{l}^{2}(\epsilon_{i})}{\pi} 
  \frac{1}{(\omega' - \omega_{1})^{2}} \nonumber\\
& + \frac{1}{\hbar} \delta_{fa} A_{ab}^{-} 
  \frac{\delta_{l}(\epsilon_{i})}{\pi} 
  \frac{ \pi \delta\left( \frac{(\omega' - \omega_{1})^{2}}{2\omega'\omega_{1}} \right) }{2\omega'\omega_{1}} 
\bigg) A_{if}^{+}
\end{align}
As the frequency of the transverse magnetic field $\omega'$ is not the same as the frequency of spin precession $\omega_1$, the above equation of $\rm{Im}\chi_A(\omega')$ is simplified to be, 
\begin{equation}
\begin{aligned}
&\mathrm{Im}\chi_{A}(\omega') \nonumber \\
&= \hbar \omega_{1} \sum_{f,i,a,b} \eta(\epsilon_{f}) \bigg(
\frac{1}{\hbar} \frac{\delta_{l}^{2}(\epsilon_{f})}{\pi} 
  A_{ab}^{-} \delta_{bi} \frac{1}{(\omega' - \omega_{1})^{2}} \\
&\qquad\qquad + \frac{1}{\hbar}  \frac{\delta_{l}^{2}(\epsilon_{i})}{\pi}  A_{ab}^{-} \delta_{fa} 
  \frac{1}{(\omega' - \omega_{1})^{2}} 
\bigg).
\end{aligned}
\label{eq:39}
\end{equation}
In natural units, 
the Gilbert damping factor is 
\begin{equation}
\begin{aligned}
\alpha= \frac{1}{2\left\langle S_{z} \right\rangle} 
\sum_{\omega_{f},l,m} \eta(\omega_{f}) \bigg(
& - \frac{\xi^{2} m^{2} m_q \omega_{f} \sigma_{l}(\omega_{f})}
       {2\pi^{2}(2l + 1)(\omega' - \omega_{1})^{2}} \\
& - \frac{\xi^{2} m^{2} m_q \omega_{f} \sigma_{l}(\omega_{f} - \omega_1)}
       {2\pi^{2}(2l + 1)(\omega' - \omega_{1})^{2}} 
\bigg)
\end{aligned}
\end{equation}
where $m$ is the magnetic angular momentum, $m_q$ is the mass of the heavy quark.

\subsection{Gilbert damping factor with two-body scatterings}
In this section, we extend the case of particle scatterings with an external spherical potential to the realistic case with two-body interactions and spin-magnetic field interactions.  The Hamiltonian of two particles in the collision process is,
\begin{equation}
H_{ee}^{(2)}=-\frac{\hbar^{2}\nabla_{r_{1}}^{2}}{m_q}-\frac{\hbar^{2}\nabla_{r_{2}}^{2}}{2m_q}+V(\left|\bf r_{1}-\bf r_{2}\right|)-\gamma  \mathbf S^{(2)}_{tot} \cdot \mathbf  B
\end{equation}
where $m_q$ is the mass of the fermion. $r_1$ and $r_2$ are the coordinates of two fermions respectively. $V(|{\bf r}_1-{\bf r}_2|)$ is the potential between two fermions.  $\gamma  \mathbf S^{(2)}_{tot} \cdot \mathbf  B $ caused the Zeeman energy level splitting of two particles. 
 
With two-body scatterings in the system, the Hamiltonian is expressed in terms of the center of mass coordinate R, relative coordinate r, and the reduced mass $m_r$, 
\begin{equation}
H^{(2)}_{ee}=-\frac{\hbar^{2}\nabla_{R}^{2}}{2(2m_q)}-\frac{\hbar^{2}\nabla_{r}^{2}}{2m_r}+V(\left|r\right|)-\gamma  \mathbf S^{(2)}_{tot} \cdot \mathbf  B.
\end{equation}
with the eigenstate defined as 
$\psi_{s_1,s_2}(R,r)=\exp(-i\mathbf k_{tot}\mathbf R)\Phi_{s_1,s_2}(r)$, and $\Phi_{s_1,s_2}(r)$ is the eigenstate of $H=-\frac{\hbar^2 \bigtriangledown_r^2}{2m_r}+V(|r|)-\gamma  \mathbf S^{(2)}_{tot} \cdot \mathbf  B$, $s_1,s_2$ represents the spin index of two particles. Here $\mathbf P=\hbar \mathbf k_{tot}$ represents the total momentum of the two particles.
The total spin $\mathbf S^{(2)}_{tot}$ of the two-particle system in the z direction is $0,\pm 1$. Choose the Pauli matrices of $S^{(2)}_{z}$ as
\begin{equation}
\begin{aligned}
S^{(2)}_{z} &= 
\begin{pmatrix}
1 & 0 & 0 \\
0 & 0 & 0 \\
0 & 0 & -1
\end{pmatrix}
\end{aligned}
\end{equation}
and the two-body operator $A^{(2)-}=[S^{(2)-},H_{so}^{(2)}]$ becomes
\begin{align}
A^{(2)-} &= {\xi}
\begin{pmatrix}
- L^{-} & 0 & 0 \\
\sqrt{2} L^{z} & 0 & 0 \\
0 & -\sqrt{2} L^{z} & L^{-}
\end{pmatrix}, \\
A^{(2)+} &= {\xi}
\begin{pmatrix}
- L^{+} & \sqrt{2} L^{z} & 0 \\
0 & 0 & -\sqrt{2} L^{z} \\
0 & 0 & L^{+}
\end{pmatrix},
\end{align}
$\xi$  takes the form of  $\xi=\frac{1}{2 m^{2} c^{2} r} \frac{\mathrm{~d} U}{\mathrm{~d} r}$ and $U$ is potential between these two particles. Then we employ the second quantization and write $\chi_A$ in the momentum space, the corresponding matrix elements of the angular momentum and the potential becomes,
\begin{align}
\left\langle ij \right| L^{z} \left| kl \right\rangle 
&= \left\langle \mathbf{p}_{i} - \mathbf{p}_{j} \right| L^{z} \left| \mathbf{p}_{k} - \mathbf{p}_{l} \right\rangle, \\
\left\langle ij \right| V \left| kl \right\rangle 
&= \left\langle \mathbf{p}_{i} - \mathbf{p}_{j} \right| V \left| \mathbf{p}_{k} - \mathbf{p}_{l} \right\rangle.
\end{align}
$i,j,k,l$ represent momentum eigenstates of these particles. $\mathbf p_{i,j,k,l}$ are the momentum of these particles. Perform the trace in the expression of $\chi_A$, one can obtain the form 
\begin{equation}
\begin{aligned}
\chi_{A}(\omega') 
&= -\xi^{2} \left( \sum_{ijkl} F \left( L_{ijkl}^{z} L_{klij}^{z} \left( \frac{1}{\omega' - \omega_1} \right) \right. \right. \\
&\quad + \sum_{ijkli'j'k'l'} V_{iji'j'}^{\dagger} L_{i'j'k'l'}^{z} \delta_{k'l'kl} L_{klij}^{z} \left( \frac{1}{\omega' - \omega_1} \right)^{2} \\
&\quad \left. \left. - \delta_{iji'j'} L_{i'j'k'l'}^{z} V_{k'l'kl} L_{klij}^{z} \left( \frac{1}{\omega' - \omega_1} \right)^{2} \right) \right),
\end{aligned}
\end{equation}
and 
\begin{equation}
\begin{aligned}
F\equiv 
&\left( f(\epsilon_{i}) - f(\epsilon_{k}) \right) f(\epsilon_{l}) f(\epsilon_{j}) 
+ \left( f(\epsilon_{j}) - f(\epsilon_{l}) \right) f(\epsilon_{i}) f(\epsilon_{k}) 
 \\&+ f(\epsilon_{k}) f(\epsilon_{l}) - f(\epsilon_{i}) f(\epsilon_{j}) 
\end{aligned}
\end{equation}
is calculated through the creation and annihilation operators in the same manner as Eq. \ref{eq:22}.

With $\mathbf{p}_{2}=\mathbf{p}_{k}-\mathbf{p}_{l},\,\mathbf{p}_{1}=\mathbf{p}_{i}-\mathbf{p}_{j}$,
\begin{equation}
\begin{aligned}
&\sum_{ijkli'j'k'l'} V_{iji'j'}^{\dagger} L_{i'j'k'l'}^{z} \delta_{k'l'kl} L_{klij}^{z} F \\
&= \sum_{ijkl} \left[ -\hbar^{2} p_{1x} p_{2x} \left( \frac{\partial}{\partial p_{1y}} \frac{\partial}{\partial p_{2y}} V_{\mathbf{p}_{1}\mathbf{p}_{2}}^{\dagger} F \right) \right. \\
&\quad + \hbar^{2} p_{1y} p_{2x} \left( \frac{\partial}{\partial p_{1x}} \frac{\partial}{\partial p_{2y}} V_{\mathbf{p}_{1}\mathbf{p}_{2}}^{\dagger} F \right) \\
&\quad + \hbar^{2} p_{1x} p_{2y} \left( \frac{\partial}{\partial p_{1y}} \frac{\partial}{\partial p_{2x}} V_{\mathbf{p}_{1}\mathbf{p}_{2}}^{\dagger} F \right) \\
&\quad \left. - \hbar^{2} p_{1y} p_{2y} \left( \frac{\partial}{\partial p_{1x}} \frac{\partial}{\partial p_{2x}} V_{\mathbf{p}_{1}\mathbf{p}_{2}}^{\dagger} F \right) \vphantom{\frac{\partial}{\partial p_{1y}}} \right]
\end{aligned}
\label{eq-59-term1}
\end{equation}
This term is non-zero only with the condition ${\bf p}_1={\bf p}_2$. The transition element $V_{{\bf p}_1{\bf p}_2}$ is, 
\begin{equation}
\begin{aligned}V_{\mathbf{p}_{1}\mathbf{p}_{2}} & =\frac{1}{2}\frac{2m_q}{\hbar^{2}}\frac{1}{iq}\frac{2\pi\hbar^{2}}{m_qL_b^{3}}\int_{0}^{\infty}\frac{r^{2}}{r}V(r)\left(e^{iqr}-e^{-iqr}\right)dr\\
 & =\frac{4\pi}{qL_b^{3}}\int_{0}^{\infty}rV(r)\sin qrdr,
\end{aligned}
\end{equation}
$L_b$ represents the length of the box after box normalization.

According to 
the optical theorem, the scattering amplitude is connected with the total cross section via $Imf\left(\mathbf{k},\mathbf{k}\right)=\frac{k\sigma_{\text{tot }}}{4\pi}$, with $f({\bf k}', {\bf k}) $ is defined as 
\begin{equation}
f\left(\mathbf{k}^{\prime},\mathbf{k}\right)\thickapprox-\frac{m_qL_b^{3}}{2\pi\hbar^{2}}\left\langle \mathbf{k}^{\prime}|V|\mathbf{k}\right\rangle ,
\end{equation}
 and $\mathbf p_1=\hbar \mathbf k',\,\mathbf p_2=\hbar \mathbf k$. Therefore, the imaginary part of $\chi_A$ can be calculated with above optical theorem. With the note of $A_1\equiv \sum_{ijkli'j'k'l'} V_{iji'j'}^{\dagger} L_{i'j'k'l'}^{z} \delta_{k'l'kl} L_{klij}^{z} F $ and $A_2\equiv \rm{Im}\sum_{ijkli'j'k'l'}\delta_{iji'j'}L_{i'j'k'l'}^{z}V_{k'l'kl}L_{klij}^{z}F$, and the imaginary part of $\sum_{ijkl}L_{ijkl}^{z}L_{klij}^{z}F$ is zero, the final expression of $\chi_A$ and the Gilbert damping factor is simplified as
\begin{equation}
\mathrm{Im}\chi_{A}=-\xi^{2}\left(A_1\left(\frac{1}{\omega'-\omega_{1}}\right)^{2}-A_2\left(\frac{1}{\omega'-\omega_{1}}\right)^{2}\right),
\end{equation}
\begin{equation}
\label{eq-final-alpha-1}
\alpha=\frac{\xi^{2}}{2\left\langle S_{z}\right\rangle \omega_{1}}\left(A_1\left(\frac{1}{\omega'-\omega_{1}}\right)^{2}-A_2\left(\frac{1}{\omega'-\omega_{1}}\right)^{2}\right).
\end{equation}
The expression of the terms $A_1$ and $A_2$ are,
\begin{equation}
\begin{aligned}
A_1 &= \sum_{ij} \left[ -\hbar^{2} p_{x} p_{x} \left( \frac{\partial}{\partial p_{y}} \frac{\partial}{\partial p_{y}} \left( \frac{\hbar^{2} p \sigma_{\text{tot}}}{2 m_q L_b^{3}} F \right) \right) \right. \\
&\quad + \hbar^{2} p_{y} p_{x} \left( \frac{\partial}{\partial p_{x}} \frac{\partial}{\partial p_{y}} \left( \frac{\hbar^{2} p \sigma_{\text{tot}}}{2 m_q L_b^{3}} F \right) \right) \\
&\quad + \hbar^{2} p_{x} p_{y} \left( \frac{\partial}{\partial p_{y}} \frac{\partial}{\partial p_{x}} \left( \frac{\hbar^{2} p \sigma_{\text{tot}} F}{2 m_q L_b^{3}} \right) \right) \\
&\quad \left. - \hbar^{2} p_{y} p_{y} \left( \frac{\partial}{\partial p_{x}} \frac{\partial}{\partial p_{x}} \left( \frac{\hbar^{2} p \sigma_{\text{tot}} F}{2 m_q L_b^{3}} \right) \right) \vphantom{\frac{\partial}{\partial p_{y}}} \right],
\end{aligned}
\end{equation}
\begin{equation}
\begin{aligned}
A_2 &= \sum_{ij} \left[ \hbar^{2} p_{x} p_{x} \left( \frac{\partial}{\partial p_{y}} \frac{\partial}{\partial p_{y}} \left( \frac{\hbar^{2} p \sigma_{\text{tot}}}{2 m_q L_b^{3}} F \right) \right) \right. \\
&\quad - \hbar^{2} p_{y} p_{x} \left( \frac{\partial}{\partial p_{x}} \frac{\partial}{\partial p_{y}} \left( \frac{\hbar^{2} p \sigma_{\text{tot}}}{2 m_q L_b^{3}} F \right) \right) \\
&\quad - \hbar^{2} p_{x} p_{y} \left( \frac{\partial}{\partial p_{y}} \frac{\partial}{\partial p_{x}} \left( \frac{\hbar^{2} p \sigma_{\text{tot}} F}{2 m_q L_b^{3}} \right) \right) \\
&\quad \left. + \hbar^{2} p_{y} p_{y} \left( \frac{\partial}{\partial p_{x}} \frac{\partial}{\partial p_{x}} \left( \frac{\hbar^{2} p \sigma_{\text{tot}} F}{2 m_q L_b^{3}} \right) \right) \vphantom{\frac{\partial}{\partial p_{y}}} \right].
\end{aligned}
\end{equation}
The magnetic field effect is mainly encoded in the Zeeman energy in the term of $F$ in above equation. This Gilbert damping factor depends on both the spin-orbital coupling term with the strength $\xi$ and also the interactions between particles.

\section{Numerical Results}
In the hot deconfined medium, we assume that the system consists of light quarks, and the interactions of quark-quark scatterings are dominated by the color-screened Coulumb potential,
\begin{equation}
V(r)=-\frac{\alpha_{c}}{r}\exp\left(-m_D r\right),
\end{equation}
where $m_{D}=T\sqrt{\frac{4\pi N_{c}}{3}\alpha_c\left(1+\frac{N_{f}}{6}\right)}$~\cite{Wen:2022yjx} is the Debye mass. Here $\alpha_c=\pi/12$ is the coupling strength and $T$ is the temperature of the medium. $N_c=3$ and $N_f=3$ are the number of colors and flavors, respectively. 
In semi-central nuclear collisions, both hot QCD medium and the strong magnetic field are generated. The magnitude of the magnetic field is taken as $eB_{ ext}=70\ m_\pi^2$ according to the simulations of Pb-Pb collisions at LHC collision energies, with $m_\pi$ being the mass of the pion. 
The corresponding Gilbert damping factor can be calculated with the formula from previous sections. 

\begin{figure}[!hbt]
    \centering
\includegraphics[width=0.42\textwidth]{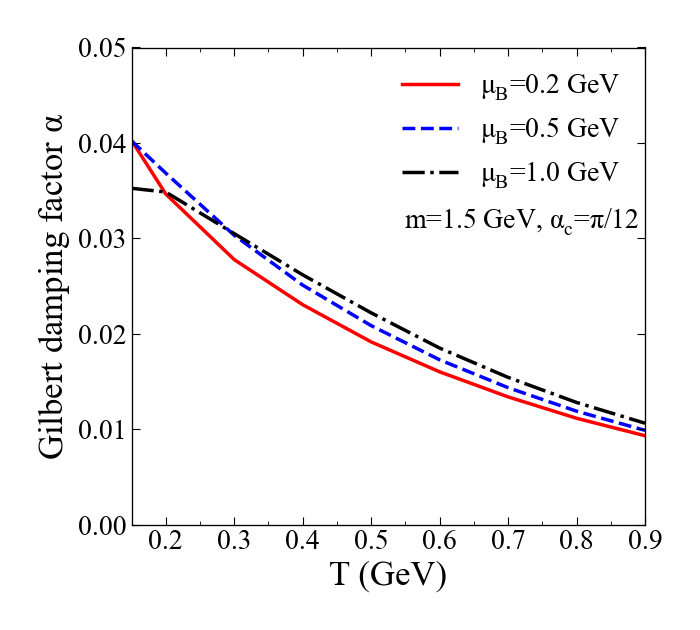}
    \caption{ The Gilbert damping factor varies with the medium temperature. The coupling parameter in the screened Coulomb potential is taken as $\alpha_{c}=\pi$/12. Heavy quark mass is taken as the charm quark mass $m_q = 1.5$ GeV. Three lines (solid, dashed, dotted-dashed) represent the case with the baryon chemical potential to be $\mu_B=0.2, 0.5, 1.0$ GeV respectively in the Fermi-distribution of light quarks in the hot QCD medium. }
    \label{fig-alpha-T}
\end{figure}

We numerically calculate the Gilbert damping factor as a function of the medium temperature in Fig.\ref{fig-alpha-T}. The baryon chemical potential which is employed in the Fermi distribution of light quarks consisting, are taken as $\mu_B=(0.2, 0.5, 1.0)$ GeV, respectively. As see in Fig.\ref{fig-alpha-T}, with a fixed external magnetic field, hotter QCD medium gives strong random scatterings on heavy quarks, which reduce the polarization rate of heavy quarks. This temperature dependence in the Gilbert damping factor is also consistent with the result in the reference~\cite{zhao2016experimental,usami2021temperature,PhysRevB.103.L220403}. Three kinds of baryon chemical potential are employed in Fig.\ref{fig-alpha-T}. As the baryon chemical potential $\mu_B$ slightly changes the densities of light quarks, which gives limited contribution in the Gilert damping factor. We approximately use the geometric mean of the upper and lower bounds of  $|\omega'-\omega_1| $ in Appendix \ref{lab-sec-chi-pm}  for numerical calculations.  

\begin{figure}[!hbt]
    \centering
\includegraphics[width=0.42\textwidth]{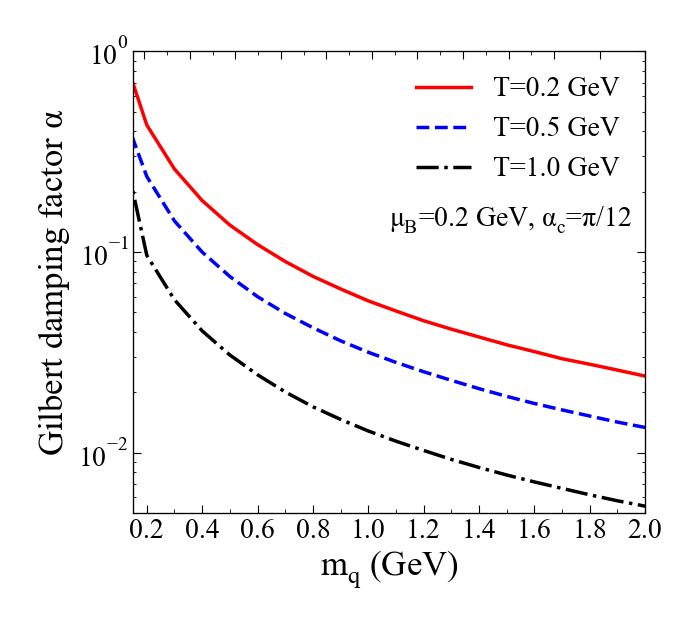}
    \caption{ The Gilbert damping factor changes with the mass of heavy quark. The coupling parameter in the Cornell potential is taken $\alpha_{c}=\pi/12$. The magnitude of the magnetic field is taken as $eB_{ ext}=70\ m_\pi^2$. Baryon chemical potential is fixed to be $\mu_B=0.2$ GeV. Three temperatures $T=(0.2, 0.5, 1.0)$ GeV are employed. }
    \label{lab-fig-alpha-mass}
\end{figure}

As the evolution of quark spins also depends on its mass, we very the mass of heavy quarks to calculate the Gilbert damping factor in Fig.\ref{lab-fig-alpha-mass}. Three values of temperature (0.2, 0.5, 1.0) GeV are considered in Fig.\ref{lab-fig-alpha-mass}. As the figure shows, the Gilbert damping factor decreases evidently with the increasing mass of heavy quarks. In the case of charm quarks, the value of Gilbert damping factor is smaller than 0.1 in the temperature regions available in RHIC and LHC nuclear collisions. 

The reliability of the formula $\alpha$ Eq.(\ref{eq-final-alpha-1}) is also discussed. 
In the derivation of $\chi_{-+}(\omega')$, the approximation of $|\omega'-\omega_1|>> |\omega_2|$ has been employed. Besides, the z-component of the quark spin vector is taken as $S_z\approx 1/2$. Therefore, the formula of the Gilbert damping factor obtained here is valid in the case of nearly complete spin polarization and the damping rate is not significant. One can also prove that with the condition of $\omega_1>>|\omega'-\omega_1|>>| \omega_2|$, our results of $\chi_{-+}({\bf q}, \omega')$ is equivalent to the case in other reference~\cite{PhysRevB.75.094416}. For the detailed proof, please see the Appendix~\ref{lab-sec-chi-pm}.

\section{summary}
In this work, we study the polarization rate of heavy quark spins in both the magnetic field and the hot QCD medium. In the presence of only a magnetic field, the spin of the heavy quark undergoes precession around the direction of the magnetic field, with a real precession frequency. When considering the combined effects of spin-magnetic field interactions and spin-spin interactions, the spin of the heavy quark becomes polarized, and the precession frequency becomes complex. This phenomenon has been widely studied using the Landau-Lifshitz-Gilbert (LLG) equation in condensed matter physics. We establish a connection between the Gilbert damping factor in the LLG equation and the complex frequency of spin waves. Furthermore, by employing the Heisenberg equation and linear response theory, we relate the transverse magnetic susceptibility of the fermionic system to the system’s Hamiltonian. Using the optical theorem, we express the scattering cross-sections in terms of the interaction terms in the Hamiltonian. Finally, we examine the two-body interactions in the medium and derive a formula for the Gilbert damping factor, 
$\alpha$, which depends on the scattering cross-sections between light quarks and the strength of the magnetic field. The temperature and mass dependence of 
$\alpha$ are also investigated. This work contributes to the understanding of spin dynamics for quarks and heavy quarkonium in the hot QCD medium.

\vspace{1cm}
{\bf Acknowledgment:} Baoyi Chen appreciates inspiring discussions with Prof. Pengfei Zhuang at the beginning of this work. This work is supported by the National Natural
Science Foundation of China (NSFC) under Grant No.
12175165.

\appendix

\section{The relationship between the damping factor and the
complex frequency}\label{sec:appendix-w2}

To calculate the relationship between the imaginary part of the frequency
in Eq. \ref{eq-s-wave} and the Gilbert damping factor, we substitute
the form of the spin from Eq. \ref{eq-s-wave} into Eq. \ref{eq-llg}.
We will obtain:

\begin{align}
(-i\omega ce^{i(qx-\omega t)},&-i\omega de^{i(qx-\omega t)},0)\nonumber \\& =(ce^{i(qx-\omega t)},de^{i(qx-\omega t)},1)\nonumber\\&\quad \times\left(i\omega\alpha ce^{i(qx-\omega t)}\text{,}\right.\nonumber \\
 &\quad \left.i\omega\alpha de^{i(qx-\omega t)},|\gamma| B_{ext}\right).\tag{A-1}
\end{align}

Since the variation in the \ensuremath{z} direction is very small,
we only need to consider the relationship in the \ensuremath{x} and
\ensuremath{y} directions. This way, we can obtain two equations:
\begin{equation}
-i\omega ce^{i(qx-\omega t)}=|\gamma| B_{ext}de^{i(qx-\omega t)}-i\omega\alpha de^{i(qx-\omega t)},\tag{A-2}\label{eq:52}
\end{equation}

\begin{equation}
-i\omega de^{i(qx-\omega t)}=-|\gamma| B_{ext}ce^{i(qx-\omega t)}+i\omega\alpha ce^{i(qx-\omega t)}.\tag{A-3}\label{eq:53}
\end{equation}

Because $\omega_{1}=|\gamma|B_{ext}$, and we set: $\lambda_1=\omega,$ $\lambda_2=\omega_{1}-i\omega\alpha$
, Eq. \ref{eq:52} and \ref{eq:53} can be written as: 
\begin{equation}
-\mathrm{i}\lambda_1c -\lambda_2 d=0,\tag{A-4}
\end{equation}
\begin{equation}
\lambda_2 c-\mathrm{i}\lambda_1 d=0.\tag{A-5}
\end{equation}

Through these two equations, we can obtain: $\text{\ensuremath{\lambda_1^{2}=\lambda_2^{2}}}$.
Here, if we set $\lambda_1=-\lambda_2$, we find that this requires $\omega_{1}$
to be a complex number. So, we set $\lambda_1=\lambda_2$. This also means
that: $\omega=\omega_{1}-i\omega\alpha$, and then we will have: $\omega_{2}=\alpha\omega_{1}/(1+\alpha^{2})$.
Since the value of $\alpha$ is generally very small, we ignore the
$\alpha^{2}$ terms in the solution. Finally, we obtain: 
\begin{equation}
\text{\ensuremath{\alpha=\omega_{2}/\omega_{1}}}.\tag{A-6}
\end{equation}
\section{The calculation of spin-orbit coupling constant \label{spin-orbit}}
In Section \ref{subsec:The-method-to},  we present the specific form of the spin-orbit coupling term, which includes a coefficient $\xi$. Next, we will conduct a detailed analysis of $\xi$. 
First, we present the Dirac equation for the electromagnetic field in the Dirac representation: 
\begin{equation}
i\gamma_{\mu}\left(\partial_{\mu}-\frac{\mathrm{i} e}{\hbar c} A_{\mu}\right) \psi-m_q \psi=0.\tag{B-1}
\end{equation}
And we set: $\psi=\binom{\varphi}{\chi}, \,\gamma_{i}=\sigma_{2} \otimes \sigma_{i},\, \gamma_{4}=\sigma_{3} \otimes \sigma_{0},  \,A_{4}=\mathrm{i} V,\, x_{4}=\mathrm{i} c t$. $\sigma _i$ is the component of the Pauli matrices, $V$ is the electric potential. $\varphi,\,\chi$ are two simple spinors of the Dirac equation. $A_\mu$ is four-vector potential. 
 Then we have: 

\begin{equation}
\label{b-2}
\left(\mathrm{i} \hbar \frac{\partial}{\partial t}-e V-m_{0} c^{2}\right) \varphi-c \sigma \cdot\left(\boldsymbol{p}-\frac{e}{c} \boldsymbol{A}\right) \chi=0,\tag{B-2}
\end{equation}
 and 
 \begin{equation}
 \label{b-3}
\left(\mathrm{i} \hbar \frac{\partial}{\partial t}-e V+m_{0} c^{2}\right) \chi-c \boldsymbol{\sigma} \cdot\left(\boldsymbol{p}-\frac{e}{c} \boldsymbol{A}\right) \varphi=0.\tag{B-3}
\end{equation}
We perform the transformation to subtract the rest energy: $\varphi(\boldsymbol{x}, t)=\mathrm{e}^{-\frac{\mathrm{i}}{\hbar} m_{0} c^{2} t} \varphi^{(\text {sub })}(\boldsymbol{x}, t), \quad \chi(\boldsymbol{x}, t)=\mathrm{e}^{-\frac{\mathrm{i}}{\hbar} m_{0} c^{2} t} \chi^{(\text {sub })}(\boldsymbol{x}, t)$, and simultaneously solve the Eq.  \ref{b-2}, Eq. \ref{b-3} :
\begin{equation}
\begin{aligned}
\left( \mathrm{i} \hbar \frac{\partial}{\partial t} - e V \right) \varphi^{(\mathrm{sub})} 
&- c \sigma \cdot \left( p - \frac{e}{c} \boldsymbol{A} \right) \frac{1}{\left( \mathrm{i} \hbar \frac{\partial}{\partial t} - e V + 2 m_{0} c^{2} \right)} \\& \times c \sigma \cdot \left( p - \frac{e}{c} \boldsymbol{A} \right) \varphi^{(\mathrm{sub})} = 0.
\end{aligned}
\tag{B-4}
\end{equation}
Then we apply the non-relativistic approximation to this equation. The first-order approximation yields\cite{thaller2005advanced}: 
\begin{equation}
 H'=\frac{\boldsymbol{p}^{2}}{2 m_{0}}+e V-\left(\mu_{L}+\mu_{S}\right) \cdot \boldsymbol{B}+\frac{e^{2} \boldsymbol{A}^{2}}{2 m_{0} c^{2}}.\tag{B-5}
\end{equation}
$\mu_L,\, \mu_S$ are orbital magnetic moment and spin magnetic moment. Next, we perform the second-order approximation:

\begin{equation}
\begin{aligned}
 H''=& \frac{1}{2m} \left[ \boldsymbol{\sigma} \cdot \left( \boldsymbol{p} - \frac{\boldsymbol{e}}{\boldsymbol{c}} \boldsymbol{A} \right) \right]^{2} + e V - \frac{\boldsymbol{p}^{4}}{8 m_{0}^{3} c^{2}}\\& + \frac{\hbar e}{4 m_{0}^{2} c^{2}} \boldsymbol{\sigma} \cdot [ (\nabla V) \times \boldsymbol{p} ] 
 + \frac{\hbar^{2} \boldsymbol{e}}{8 m_{0}^{2} c^{2}} \Delta V.
\end{aligned}
\tag{B-6}
\end{equation}
 
For the fourth term of this equation, in a stable field, it can be directly written in the familiar form of spin-orbit coupling:
\begin{equation}
\frac{e \hbar}{4 m_{0}^{2} c^{2}} \sigma \cdot((\nabla V) \times \boldsymbol{p})=\frac{1}{2 m_{0}^{2} c^{2} r} \frac{\mathrm{~d} U}{\mathrm{~d} r} \boldsymbol{S} \cdot \boldsymbol{L}.\tag{B-7}
\end{equation}
This term is actually the interaction energy between the electron's magnetic moment and the magnetic field generated by the electron's spatial orbital motion. The coefficient $\xi=\frac{1}{2 m_{0}^{2} c^{2} r} \frac{\mathrm{~d} U}{\mathrm{~d} r}$ can describe the strength of spin-orbit coupling.

\section{ The transverse susceptibility and the Hamiltonian of the fermionic system\label{sec:appendix-b.-Calculate}}

In Section \ref{subsec:Calculate-the-Gilbert-damping}, our second method
for calculating the transverse susceptibility is in the Heisenberg
picture, using the system's Hamiltonian to express the time evolution
of the spin operator. Now, we can proceed with further calculations
based on Eq. \ref{eq-chi-simp2}. Since the electron interaction and the
potential energy provided by the electric field do not directly affect the
spin, Eq. \ref{eq-chi-simp2} can be written as: 
\begin{equation}
\begin{aligned}
\chi_{-+}(\boldsymbol{q}, \omega') 
&= \frac{1}{\hbar \omega'_{-}} \bigg(
\left\langle 
\left[ S_{\boldsymbol{q}}^{-}, S_{-\boldsymbol{q}}^{+} \right] 
\right\rangle 
\\
&\quad + \frac{1}{i\hbar} \int_{0}^{\infty} dt\,
\left\langle 
\left[ 
\left[ S_{\boldsymbol{q}}^{-}(t), H \right], 
S_{-\boldsymbol{q}}^{+} 
\right] 
\right\rangle 
e^{-i \omega'_{-} t} 
\bigg)
\\[1em]
&= \frac{1}{\hbar \omega'_{-}} \bigg(
-2\hbar \left\langle S_{z} \right\rangle 
 + \frac{1}{i\hbar} \int_{0}^{\infty} dt\, \big\langle 
\left[ 
C_{\boldsymbol{q}}^{-}(t), S_{-\boldsymbol{q}}^{+} 
\right] 
\\
&\quad + \left[ 
\left[ S_{\boldsymbol{q}}^{-}(t), H_{Z} \right], 
S_{-\boldsymbol{q}}^{+} 
\right] 
\big\rangle 
e^{-i \omega'_{-} t} 
\bigg)
\end{aligned}
\tag{C-1}
\label{eq:57}
\end{equation}

In the above equation, $C_{\boldsymbol{q}}^{-}=\left[S_{\boldsymbol{q}}^{-},H_{{k}}+H_{{SO}}\right]$.
This equation consists of three terms, so it can be split into:$\chi_{-+}(\boldsymbol{q},\omega')=\frac{-1}{\hbar\omega'_{-}}\left(2\hbar\left\langle S_{z}\right\rangle +\chi_{2}+\chi_{3}\right)$.
Similar to equation Eq. \ref{eq:57}, $\chi_{2}$ can be calculated
as follows: 
\begin{equation}
\begin{aligned}
\chi_{2}(\boldsymbol{q}, \omega') 
&= \frac{1}{\hbar \omega'_{-}} \bigg( 
\left\langle 
\left[ C_{\boldsymbol{q}}^{-}, S_{-\boldsymbol{q}}^{+} \right] 
\right\rangle 
\\
&\quad + \frac{1}{i \hbar} \int_{0}^{\infty} dt\, 
\left\langle 
\left[ 
\left[ C_{\boldsymbol{q}}^{-}, H \right], 
S_{-\boldsymbol{q}}^{+} 
\right] 
\right\rangle 
e^{-i \omega'_{-} t} 
\bigg) 
\\[1em]
&= \frac{1}{\hbar \omega'_{-}} \bigg( 
\left\langle 
\left[ C_{\boldsymbol{q}}^{-}, S_{-\boldsymbol{q}}^{+} \right] 
\right\rangle 
\\
&\quad + \frac{1}{i \hbar} \int_{0}^{\infty} dt\, 
\left\langle 
\left[ 
C_{\boldsymbol{q}}^{-}, 
\left[ H, S_{-\boldsymbol{q}}^{+} \right] 
\right] 
\right\rangle 
e^{-i \omega'_{-} t} 
\bigg)
\end{aligned}
\tag{C-2}
\end{equation}

The second step of the calculation uses the following relation: $\left[\left[C_{\boldsymbol{q}}^{-}(t),H\right],S_{-\boldsymbol{q}}^{+}\right]=\left[C_{\boldsymbol{q}}^{-}(t),\left[H,S_{-\boldsymbol{q}}^{+}\right]\right]-\left[\left[C_{\boldsymbol{q}}^{-}(t),S_{-\boldsymbol{q}}^{+}\right],H\right]$
and $\langle[X,H]\rangle=0$ ~\cite{PhysRevB.90.014420}. And because:
\begin{equation}
\begin{aligned}
\left[ H, S_{-\boldsymbol{q}}^{+} \right] 
&= \left[ H_{\text{k}} + H_{\text{ee}} + H_{\text{ext}} + H_{Z} + H_{\text{SO}}, S_{-\boldsymbol{q}}^{+} \right] \\
&= \left[ H_{\text{k}} + H_{Z} + H_{\text{SO}}, S_{-\boldsymbol{q}}^{+} \right] \\
&= \left[ H_{\text{k}} + H_{\text{SO}}, S_{-\boldsymbol{q}}^{+} \right] 
  + \left[ H_{Z}, S_{-\boldsymbol{q}}^{+} \right] \\
&= C_{\boldsymbol{q}}^{+} - 2 \mu_{\mathrm{Bo}} B_{\mathrm{ext}} S_{-\boldsymbol{q}}^{+}
\end{aligned}
\tag{C-3}
\end{equation}

we can further calculate $\chi_{2}(\boldsymbol{q},\omega')$:

\begin{equation}
\begin{aligned}
\chi_{2}(\boldsymbol{q}, \omega') 
&= \frac{1}{\hbar \omega'_{-}} \Bigg( 
    \left\langle \left[ C_{\boldsymbol{q}}^{-}, S_{-\boldsymbol{q}}^{+} \right] \right\rangle 
    \\&\quad+ \frac{1}{i \hbar} \int_{0}^{\infty} dt\,
      \left\langle \left[ C_{\boldsymbol{q}}^{-}(t), C_{\boldsymbol{q}}^{+} \right] \right\rangle 
      e^{-i \omega'_{-} t} \\
&\quad 
    - \frac{2}{i \hbar}\mu_{Bo}  B_{\mathrm{ext}} 
    \int_{0}^{\infty} dt\,
    \left\langle \left[ C_{\boldsymbol{q}}^{-}(t), S_{-\boldsymbol{q}}^{+} \right] \right\rangle 
    e^{-i \omega'_{-} t} 
\Bigg)
\end{aligned}
\tag{C-4}
\end{equation}

If we introduce $\chi_{C}=\frac{i}{\hbar}\int_{0}^{\infty}dt\left\langle \left[C_{\boldsymbol{q}}^{-}(t),C_{\boldsymbol{q}}^{+}\right]\right\rangle e^{-i\omega'_{-}t}$,
this equation can be further simplified to: 
\begin{equation}
\chi_{2}(\boldsymbol{q},\omega')=\frac{1}{\hbar\omega'_{-}}\left(\left\langle C_{\boldsymbol{q}}^{-}(t),S_{-\boldsymbol{q}}^{+}\right\rangle -\chi_{C}+2\mu_{{Bo}}B_{{ext}}\chi_{2}\right).\tag{C-5}
\end{equation}

Finally, by introducing $\hbar\omega_{1}=2\mu_{Bo}B_{{ext}}$,
$\chi_{2}$ can be written as: 
\begin{equation}
\chi_{2}(\boldsymbol{q},\omega')=\frac{\left\langle \left[C_{\boldsymbol{q}}^{-},S_{-\boldsymbol{q}}^{+}\right]\right\rangle -\chi_{C}(\boldsymbol{q},\omega')}{\hbar\left(\omega'_{-}-\omega_{1}\right)}.\tag{C-6}\label{eq:62}
\end{equation}

For the third term in Eq. \ref{eq:57}, its calculation is relatively
straightforward: 
\begin{align}
\chi_{3}(\boldsymbol{q},\omega') & =\frac{i}{\hbar}\int_{0}^{\infty}dt\left\langle \left[\left[S_{\boldsymbol{q}}^{-}(t),H_{Z}\right],S_{-\boldsymbol{q}}^{+}\right]\right\rangle e^{-i\omega'_{-}t}\nonumber \\
 & =\frac{i}{\hbar}\int_{0}^{\infty}dt\left\langle \left[\left[S_{\boldsymbol{q}}^{-}(t),-2\mu_{{Bo}} B_{{ext}}S_{z}\right],S_{-\boldsymbol{q}}^{+}\right]\right\rangle e^{-i\omega'_{-}t}\tag{C-7}\label{eq:63}\\
 & =-\hbar\omega_{1}\chi_{-+}(\boldsymbol{q},\omega')\nonumber 
\end{align}

Combining Eq. \ref{eq:57}, \ref{eq:62}, and \ref{eq:63}, we finally
obtain the value of the susceptibility as: 
\begin{equation}
\begin{aligned}\chi_{-+}(\boldsymbol{q},\omega') & =\frac{-2\left\langle S_{z}\right\rangle }{\hbar\text{(}\omega'_{-}-\omega_{1})}-\frac{\chi_{2}}{\hbar\left(\omega'_{-}-\omega_{1}\right)}\\
 & =\frac{-2\left\langle S_{z}\right\rangle }{\hbar\text{(}\omega'_{-}-\omega_{1})}\\&+\frac{1}{\hbar^{2}\left(\omega'_{-}-\omega_{1}\right)^{2}}\left\{ \chi_{C}(\boldsymbol{q},\omega')-\left\langle \left[C_{\boldsymbol{q}}^{-},S_{-\boldsymbol{q}}^{+}\right]\right\rangle \right\} .
\end{aligned}
\tag{C-8}
\end{equation}

Since we are considering the case where $q\rightarrow0$, $C_{\boldsymbol{q}}^{-}=A_{\boldsymbol{q}}^{-}=A^{-}=\left[S^{-},H_{\mathrm{SO}}\right]$,
and the singularity of this equation is $\omega_{1}-i\omega_{2}$,
which can absorb the imaginary part of $\omega'_{-}$, this equation
can ultimately be written as: 
\begin{equation}
\begin{aligned}
\chi_{-+}(\omega')=&\frac{-2\left\langle S_{z}\right\rangle }{\hbar\text{(}\omega'-\omega_{1})}\\&+\frac{1}{\hbar^{2}\left(\omega'-\omega_{1}\right)^{2}}\left\{ \chi_{A}(\omega')-\left\langle \left[A^{-},S^{+}\right]\right\rangle \right\} .
\end{aligned}
\tag{C-9}
\end{equation}

\section{The particle number operator and the Fermi distribution
\label{lab-append-c}}
In Eq. \ref{eq:22}, we obtained the particle number operators $n_{i}$
and $n_{f}$. Now, let's calculate the expectation value of the difference
between the particle number operators. $Tr\left(\hat{\rho}\left(\hat{n_{f}}-\hat{n_{i}}\right)\right)$.
In the grand canonical ensemble: 
\begin{equation}
Tr\left(\hat{\rho}\left(\hat{n_{f}}-\hat{n_{i}}\right)\right)=Tr\left(\frac{1}{\Xi}\exp(-\alpha'\hat{N}-\beta\hat{H})\left(\hat{n_{f}}-\hat{n_{i}}\right)\right),\tag{D-1}\label{eq:66}
\end{equation}
where the grand partition function $\Xi$ represents $Tr\left(\exp(-\alpha'\hat{N}-\beta\hat{H})\right)$
and can be written as: 
\begin{equation}
\begin{aligned}\Xi & =\sum_{\left\{ n_{k}\right\} }\exp\left(-\alpha'\sum_{k}n_{k}-\beta\sum_{k}n_{k}\epsilon_{k}\right)\\
 & =\sum_{\left\{ n_{k}\right\} }\prod_{k}\exp\left(-\alpha' n_{k}-\beta n_{k}\epsilon_{k}\right)\\
 & =\prod_{k}\sum_{n_{k}}\exp\left(-\left(\alpha'+\beta\epsilon_{k}\right)n_{k}\right)\\
 & =\prod_{k}\Xi_{k}.
\end{aligned}
\tag{D-2}
\end{equation}

In the above equation, we use the occupation number of each single-particle
state to represent these states: $\left\{ n_{k}\right\} =\left\{ n_{1},n_{2},n_{3},\cdots\right\} $.
Using a similar method, we can perform the calculation for Eq. \ref{eq:66}:
\begin{equation}
\begin{aligned}
\mathrm{Tr}\Bigg(&\frac{1}{\Xi} \exp(-\alpha' \hat{N} - \beta \hat{H}) 
\left( \hat{n}_{f} - \hat{n}_{i} \right) \Bigg) \\
&= \frac{1}{\Xi} \sum_{\{n_k\}} 
\exp(-\alpha' N - \beta E_{\{n_k\}}) (n_f - n_i) \\
&= \frac{1}{\Xi} \sum_{\{n_k\}} 
\exp\left( -\alpha' \sum_k n_k - \beta \sum_k n_k \epsilon_k \right) 
(n_f - n_i) \\
&= \frac{1}{\Xi} \sum_{\{n_k\}} 
\prod_k \exp(-\alpha' n_k - \beta n_k \epsilon_k) 
(n_f - n_i)
\end{aligned}
\tag{D-3}
\label{eq:68}
\end{equation}

We consider the first part of the equation $\frac{1}{\Xi}\sum_{\left\{ n_{k}\right\} }\prod_{k}\exp(-\alpha' n_{k}-\beta n_{k}\epsilon_{k})n_{f}$,
and for clarity, we define a function $g(k)$ such that: 
\begin{equation}
g(k)=\begin{cases}
\exp(-\alpha' n_{k}-\beta n_{k}\epsilon_{k}), & k\neq f\\
\exp(-\alpha' n_{f}-\beta n_{f}\epsilon_{f})n_{f}. & k=f
\end{cases}\tag{D-4}
\end{equation}

Then, we can write the first part of Eq. \ref{eq:68} as: 
\begin{equation}
\frac{1}{\Xi}\sum_{\left\{ n_{k}\right\} }\prod_{k}g(k)=\frac{1}{\Xi}\prod_{k}\sum_{n_{k}}g(k)=\prod_{k}\frac{\sum_{n_{k}}g(k)}{\Xi_{k}},\tag{D-5}
\end{equation}

and it can be calculated as:

\begin{align}
\prod_{k}\frac{\sum_{n_{k}}g(k)}{\Xi_{k}} & =\frac{\sum_{n_{1}}\exp(-\alpha' n_{1}-\beta n_{1}\epsilon_{1})}{\sum_{n_{1}}\exp(-\alpha' n_{1}-\beta n_{1}\epsilon_{1})}\nonumber\\&\ldots\frac{\sum_{n_{f}}\exp(-\alpha' n_{f}-\beta n_{f}\epsilon_{f})n_{f}}{\sum_{n_{f}}\exp(-\alpha' n_{f}-\beta n_{f}\epsilon_{f})}\cdots\nonumber \\
 & =\frac{\sum_{n_{f}}\exp(-\alpha' n_{f}-\beta n_{f}\epsilon_{f})n_{f}}{\sum_{n_{f}}\exp(-\alpha' n_{f}-\beta n_{f}\epsilon_{f})}\tag{D-6}\\
 & =-\frac{\partial}{\partial\alpha'}\ln\Xi_{f}.\nonumber 
\end{align}

For fermions: $n_{k}=0,1,~\Xi_{k}=1+\exp(-\alpha-\beta\epsilon_{k})$.
Thus, in the end, we calculate that the first half of Eq. \ref{eq:68}
is: 
\begin{equation}
\begin{aligned}
\frac{1}{\Xi} \sum_{\{n_k\}} 
\prod_k \exp(-\alpha' n_k - \beta n_k \epsilon_k) n_f 
&= \frac{1}{\exp\left(\alpha' + \beta \epsilon_f\right) + 1} \\
&= \frac{1}{\exp\left(\frac{\epsilon_f - \mu_B}{k_bT}\right) + 1}.
\end{aligned}
\tag{D-7}
\end{equation}

Using the same method, we can calculate the second half of Eq. \ref{eq:68}:
\begin{equation}
\frac{1}{\Xi}\sum_{\left\{ n_{k}\right\} }\prod_{k}\exp(-\alpha' n_{k}-\beta n_{k}\epsilon_{k})n_{i}=\frac{1}{\exp\left(\frac{\epsilon_{i}-\mu_B}{k_bT}\right)+1}.\tag{D-8}
\end{equation}
Finally we have: 
\begin{equation}
\begin{aligned}
\mathrm{Tr}\Bigg(&\frac{1}{\Xi} \exp(-\alpha' \hat{N} - \beta \hat{H}) 
\left( \hat{n}_{f} - \hat{n}_{i} \right) \Bigg) 
= f(\epsilon_f)-f(\epsilon_i).
\end{aligned}
\tag{D-9}
\end{equation}
$f(\epsilon)$ is fermion distribution.

\section{Derivation of $\chi_{-+}$\label{lab-sec-chi-pm}}
From this paper~\cite{PhysRevB.75.094416}, we obtained
an alternative method for calculating the transverse magnetic susceptibility.
First, we add an additional term to the Hamiltonian: 
\begin{equation}
H_{add}=-\gamma\mathbf{h\cdot S},
\tag{E-1}
\end{equation}

where $\mathbf{h}$ represents a transverse infinitesimal magnetic
field. Next, we use the LLG equation: 
\begin{equation}
\begin{aligned}
\left( \frac{\partial s_{x}}{\partial t}, \frac{\partial s_{y}}{\partial t}, 0 \right) 
= \left( s_{x}, s_{y}, 1 \right) \times 
&\left( -\alpha \frac{\partial s_{x}}{\partial t} - |\gamma| h_{x}, \right. \\
&\left. -\alpha \frac{\partial s_{y}}{\partial t} - |\gamma |h_{y},\ |\gamma |B_{ext} \vphantom{\frac{\partial}{\partial t}} \right),
\end{aligned}
\tag{E-2}
\end{equation}

In the x-direction and y-direction, we obtain: 
\begin{equation}
\frac{\partial s_{x}}{\partial t}=s_{y}\left(|\gamma| B_{ext}\right)-\left(-\alpha\frac{\partial s_{y}}{\partial t}-|\gamma| h_{y}\right),
\tag{E-3}
\end{equation}

\begin{equation}
\frac{\partial s_{y}}{\partial t}=-s_{x}\left(|\gamma| B_{ext}\right)+\left(-\alpha\frac{\partial s_{x}}{\partial t}-|\gamma |h_{x}\right).
\tag{E-4}
\end{equation}

After performing a Fourier transform, in the frequency domain, we
obtain: 
\begin{equation}
s_{-}=-\frac{|\gamma| h_{-}}{\omega_{1}-\omega'+\mathrm{i}\alpha\omega'}.
\tag{E-5}
\end{equation}

This leads to an alternative expression for the transverse magnetic
susceptibility: 
\begin{equation}
\chi_{-+}(\omega')=\frac{2\left\langle S_{z}\right\rangle }{\omega_{1}-\omega'+\mathrm{i}\alpha\omega'}.
\tag{E-6}
\end{equation}

It can also be written as: 
\begin{equation}
\chi_{-+}(\omega')=\frac{2\left\langle S_{z}\right\rangle }{\omega_{1}-\omega'+\mathrm{i}\alpha\left(\omega'-\omega_{1}+\omega_{1}\right)}.
\tag{E-7}
\end{equation}

Comparing with Eq. \ref{eq-chi-final-1}, we find that when $\omega$ satisfies
condition $\omega_{1}\gg|\omega'-\omega_{1}|$, the two expressions are
equal. This means that, to calculate the Gilbert damping coefficient
$\alpha$, at least the following two conditions must be satisfied:
$|\omega'-\omega_{1}|\gg|\omega_{2}|,\omega_{1}\gg|\omega'-\omega_{1}|.$

\bibliography{sub-v1}

\end{document}